%% file: poseidon.tex
\documentclass[sigconf, nonacm]{acmart}

\newcommand\vldbdoi{XX.XX/XXX.XX}

\newcommand\vldbavailabilityurl{}
\newcommand\vldbpagestyle{plain}

\usepackage{xspace}
\usepackage{balance}
\usepackage[colorinlistoftodos,prependcaption,textsize=tiny,%
	]{todonotes}
\usepackage{algorithm}
\usepackage{algpseudocode}

\newcommand{\Peight}{{Poseidon}\xspace}
\newcommand{\gui}{{gui}\xspace}
\newcommand{\APL}{{P8APL}\xspace}
\newcommand{\RELVERT}{{\sc Rel\_Vert}\xspace}
\newcommand{\RELLIT}{{\sc Rel\_Lit}\xspace}
\newcommand{\RELE}{{\sc Rel\_E}\xspace}
\newcommand{\RELEP}{{\sc Rel\_Ep}\xspace}
\newcommand{\RELVP}{{\sc Rel\_Vp}\xspace}
\newcommand{\RELMVP}{{\sc Rel\_Mvp}\xspace}
\newcommand{\RELME}{{\sc Rel\_Me}\xspace}
\newcommand{\RELMEP}{{\sc Rel\_Mep}\xspace}
\newcommand{\RELLME}{{\sc Rel\_Lme}\xspace}
\newcommand{\RELLMEP}{{\sc Rel\_Lmep}\xspace}
\newcommand{\RELRME}{{\sc Rel\_Rme}\xspace}
\newcommand{\RELRMEP}{{\sc Rel\_Rmep}\xspace}
\newcommand{\RELVSS}{{\sc Rel\_Vss}\xspace}

\begin{document}
\title{Poseidon: A OneGraph Engine}
\author{Brad Bebee}
\affiliation{%
  \institution{Amazon Web Services}
  \city{Seattle, Washington}
  \country{USA}
}
\email{beebs@amazon.com}

\author{\"Umit V. \c{C}ataly\"urek}
\authornote{\"Umit V. \c{C}ataly\"urek is appointed both as an Amazon Scholar and as a Professor at Georgia Institute of Technology. This paper describes work performed at Amazon.}
\affiliation{%
  \institution{Amazon Web Services}
  \city{Atlanta, Georgia}
  \country{USA}}
\email{uvc@amazon.com}

\author{Olaf Hartig}
\authornote{Olaf Hartig is appointed both as an Amazon Scholar and as a Senior Associate Professor at Linköping University. This paper describes work performed at Amazon.}
\orcid{0000-0002-1741-2090}
\affiliation{%
  \institution{Amazon Web Services}
  \country{Sweden}}
\email{olaf.hartig@liu.se}

\author{Ankesh Khandelwal}
\affiliation{%
  \institution{Amazon Web Services}
  \city{Seattle, Washington}
  \country{USA}
}
\email{ankeshk@amazon.com}

\author{Simone Rondelli}
\affiliation{%
  \institution{Amazon Web Services}
  \city{Seattle, Washington}
  \country{USA}
}
\email{rondelli@amazon.com}

\author{Michael Schmidt}
\affiliation{%
  \institution{Amazon Web Services}
  \city{Seattle, Washington}
  \country{USA}
}
\email{schmdtm@amazon.com}

\author{Lefteris Sidirourgos}
\affiliation{%
  \institution{Amazon Web Services}
  \state{Germany}
}
\email{sidirol@amazon.com}
\authornote{Corresponding author}

\author{Bryan Thompson}
\affiliation{%
  \institution{Amazon Web Services}
  \city{Seattle, Washington}
  \country{USA}
}
\email{bryant@amazon.com}

\begin{abstract}
  We present the \Peight\ engine behind the Neptune Analytics graph database service.  Customers interact with \Peight using the declarative openCypher~\cite{green2018opencypher} query language, which enables requests that seamlessly combine traditional querying paradigms (such as graph pattern matching, variable length paths, aggregation) with algorithm invocations and has been syntactically extended to facilitate {\em OneGraph} interoperability, such as the disambiguation between globally unique IRIs (as exposed via RDF) vs.~local identifiers (as encountered in LPG data). \Peight supports a broad range of graph workloads, from simple transactions, to top-k beam search algorithms on dynamic graphs,
to whole graph analytics requiring multiple full passes over the data. For example, real-time fraud detection, like many other
use cases, needs to reflect current committed state of the dynamic graph. If a user's cell phone is compromised, then all newer actions by that
user become immediately suspect.  To address such dynamic graph use cases, \Peight combines state-of-the-art transaction processing with novel graph data indexing, including lock-free maintenance of adjacency
lists, secondary succinct indices, partitioned heaps for data tuple storage with uniform placement, and innovative statistics for
cost-based query optimization. The \Peight engine uses a {\em logical log} for durability, enabling rapid evolution of
in-memory data structures. Bulk data loads achieve more than 10M property values per second on many data sets while simple transactions
can execute in under 20${\mu}s$ against the storage engine.
\end{abstract}

\maketitle
\pagestyle{\vldbpagestyle}

\section{Introduction}\label{sec:intro}
\input{intro}

\section{1G Abstract Model}\label{sec:1g}
\input{1g}

\section{1G Physical Data Model}\label{sec:rel}
\input{rel}

\section{Transactions and Durability}\label{sec:storage}
\input{storage}

\section{Data Scan Operations}\label{sec:index}

\input{index}

\section{Access Patterns for 1G Data}\label{sec:apl}
\input{apl}

\section{Graph Algorithms}\label{sec:algos}
\input{algos}

\section{Query Processing}\label{sec:queries}
\input{queries}

\section{Performance Evaluation}\label{sec:perf}
\input{perf}

\section{Conclusion and Future Work}\label{sec:concl}
We have described \Peight, a novel hybrid transactional and analytic engine for graph workloads.  The \Peight\ engine has been in production since AWS re:Invent 2023, powering the Neptune Analytics service.  Neptune Analytics customers have created thousands of graphs since launch. Recently, a new group of customers is using Neptune Analytics to support
their GraphRAG
\footnote{https://aws.amazon.com/about-aws/whats-new/2025/03/amazon-bedrock-knowledge-bases-graphrag-generally-available/}
workloads.  Overall, we have achieved our initial design objectives, delivering a high performance graph engine to Amazon customers.  Customers 
are able to use graph traversals with HPC performance against live data while multi-pass graph analytics transparently use HPC indexing strategies when the cost
of building the index is justified.  Query workloads automatically switch between relation scans and index scans based on the selectivity and the query optimizer
has access to detailed statistics and a detailed cost model of engine operations.

Decoupling the in-memory indices from the durable storage using a logical log has made it possible for us to continuously reinvent the engine, more than doubling
the load performance since the service was announced.  We discovered some performance artifacts in production (especially for \RELE) and have been able to address
those concerns by replacing the use of concurrent hash maps with simple arrays backed by virtual memory allocations.  Future work will further smooth out performance
using a hybrid indexing strategy, which also improves data density for hyper-sparse graphs, replace the remaining concurrent hash map over "O" property values with a
thread-safe ordered index; and introduce zone maps, small aggregates, and imprints~\cite{Sidirourgos13} to further accelerate relation scans and data access.

\begin{acks}
We would like to acknowledge all of the members of the Amazon Neptune team for their contributions to Neptune Analytics.
A special thanks to Tengiz Kharatishvili for his insightful contributions on transactions and durability.  
\end{acks}

\balance
\bibliographystyle{ACM-Reference-Format}
\bibliography{references,tdalab,pgabb}
\end{document}

%% file: intro.tex
Since its launch in 2017, 
thousands of customers have created tens of thousands of Amazon Neptune graph data clusters because of the choice of data models and query languages
(Labeled Property Graph (LPG)~\cite{rodriguez2010,Robinson15:GraphDBsBook2ndEd} or 
Resource Description Framework (RDF)~\cite{RDF11Concepts}), enterprise features, and the simplicity of a fully managed 
graph database experience. However, customers have
asked for LPG/RDF interoperability, higher load rates, better low latency query performance, and graph analytics support.
To address these customer needs we 
launched Neptune Analytics, a service providing graph query, graph algorithms, and graph analytics
based on a new engine,
\emph{\Peight}. The \Peight engine, based on the {\em OneGraph} (1G) data 
model~\cite{DBLP:journals/semweb/LassilaSHBBBKLL23}, directly addresses interoperability in the data between RDF and LPG
data~\cite{DBLP:conf/semweb/0002BBEENSSTVWX24} and fully supports both graph representations, including LPG concepts such
as multiple edges for the same source and target vertex, meta-properties, edge properties, etc.

With Neptune Analytics, you can get insights and find trends by processing large amounts of graph data in seconds. To analyze graph data quickly and easily, Neptune Analytics stores large graph datasets in memory. It supports a library of optimized graph analytic algorithms, low-latency graph queries, and vector search capabilities. Customers use Neptune Analytics for investigatory, exploratory, or data-science workloads that require fast iteration for data, analytical and algorithmic processing, or vector search on graph data. It complements Amazon Neptune Database, a popular managed graph database. To perform intensive analysis, you can load the data from a Neptune Database graph or snapshot into Neptune Analytics. You can also load graph data that's stored in Amazon S3.

Currently, customers interact with Neptune Analytics using the declarative openCypher~\cite{green2018opencypher}
query language, which enables requests that combine traditional querying paradigms (such as graph pattern matching, variable-length paths, aggregation) with algorithm invocations and has been syntactically extended to facilitate {\em OneGraph} interoperability, such as the disambiguation between globally unique IRIs (as exposed via RDF) vs.~local identifiers (as encountered in LPG data). \Peight supports a broad range of graph workloads, from simple transactions, to top-k beam search algorithms on dynamic graphs,
to whole graph analytics requiring multiple full passes over the data. For example, real-time fraud detection, like many other
use cases, needs to reflect current committed state of the dynamic graph. If a user's cell phone is compromised, then all newer actions by that
user become immediately suspect.  To address such dynamic graph use cases, \Peight combines state-of-the-art transaction processing with novel graph data indexing, including lock-free maintenance of adjacency
lists, succinct secondary indices, partitioned heaps for data tuple storage with uniform placement, and innovative statistics for
cost-based query optimization. The \Peight engine uses a {\em logical log} for durability, enabling rapid evolution of
in-memory data structures. Bulk data loads achieve more than
10M property values per second on many data sets while simple transactions
can execute in under 20${\mu}s$ against the storage engine.

\Peight is a main memory, high performance, low latency,
hybrid transactional and analytics (HTAP) graph storage engine. Its hybrid columnar design is heavily influenced by advances
in a)~High Performance Computing (HPC) graph algorithms, b)~Main-Memory
Databases for efficient data retrieval, and c)~Transactional Databases for supporting demanding online transactional
workloads.  

\Peight seeks to balance performance requirements for {\em graph algorithms}, as well as for {\em OLAP} and {\em OLTP} queries.
To achieve this goal, we designed and built a hybrid in-memory data storage engine with 
graph specific partitioning and indices. 
\Peight maintains the locality of specific graph access paths: the property set of a vertex is always within a
single partition, and the edge properties are {\em co-located} in the same {\em container} as the edges that they describe. On the
other hand, data about the {\em topology} of the graph (i.e., edges connecting resources) is distributed across partitions,
which ensures that we can scale graph algorithms when the input frontier becomes large. 
While the \Peight engine does well on both OLTP and OLAP workloads, OLAP workloads
can be run on separate compute to minimize the impact on transactional workloads. 

The graph data are organized as a system of partitioned relations for string data, vertex properties,
edges, edge properties, etc.  Internally, each partition maintains a
heap of narrow tuples on PAX pages~\cite{Ailamaki01}. Secondary indices are maintained
over that heap.  Graph specific indices and lock-free data structures
are used throughout to achieve peak performance against a highly
dynamic transactional graph.  The query engine and algorithm kernels 
interact with an {\em Access Pattern Language} (\APL) which decouples the
internal organization of memory and provides low level data access optimizations.
The engine is also decoupled from the durable representation using a logical log, 
thus enabling very rapid evolution of the different components.
%

The contributions of this publication are:
\begin{enumerate}
\item The 1G abstract model;
\item A physical realization of the 1G abstract model, including the system of partitioned relations used to model a 1G graph in \Peight, the physical organization of tuples on PAX pages in memory, the use of secondary indices, etc.;
\item A discussion of transactions and durability, including the use of fast mvcc and how using a logical log accelerates innovation;
\item A discussion of \APL and scans, including the description of the access pattern language which provides a separation of concerns encapsulating access to the physical organization of the graph in memory, and how the \APL selects the most appropriate low level kernels for each access pattern expression;
\item A discussion of graph algorithms in \Peight, including how algorithms can run against live data or against static views from a point in time, how algorithm kernels are written and reused internally by the engine, dynamic switching between index-driven and scan-driven, and how algorithms are parallelized; and
\item We sketch key aspects of the query language and runtime query processing and algorithm integration, and describe design principles behind the interaction of the cost-based query optimizer with storage-level APIs for cardinality estimation and higher-level statistics.
\end{enumerate}

%% file: 1g.tex
	%
	%
	The Neptune graph database service supports both the graph data model of RDF~\cite{RDF11Concepts}, with its query language SPARQL~\cite{Harris+Seaborne-SPARQLSpec2013}, and LPGs~\cite{rodriguez2010,Robinson15:GraphDBsBook2ndEd}, with the query languages Gremlin~\cite{rodriguez2015} and openCypher~\cite{DBLP:conf/edbt/GreenJKLPS18}. Yet, customers have
	%
to choose one of these models when creating a new database instance and could not ``cross-use'' these technologies~(e.g., 
 	querying LPGs using SPARQL, or RDF using openCypher).
We have observed that this limitation can cause confusion (especially, for users who are new to graphs) and making the choice is not trivial as it requires considering data modeling aspects, query language features, adequacy for current and future use cases, and developer knowledge%
.
To remove these obstacles 
	and achieve interoperability between these technologies we are pursuing the OneGraph~(1G)~project%
~\cite{DBLP:journals/semweb/LassilaSHBBBKLL23}.

An initial result of this effort---available to users of Neptune Analytics, on top of \Peight---is the option to load LPG and RDF data into a single, connected graph and to query this graph using openCypher~\cite{DBLP:conf/semweb/0002BBEENSSTVWX24}. The conceptual basis of this functionality is an abstract data model, called the \emph{1G model}, which combines features from both LPGs and RDF and can be used as a foundation for switching seamlessly between an RDF and an LPG view of the data.

While presenting the formal definition of this model and relevant data and query language mappings is beyond the scope of this paper,
we briefly summarize the main concepts of the model: A graph dataset in the 1G model consists of a set of so-called \emph{1G elements} that have an identity and can assume different roles within the dataset. In particular, every 1G element may represent a node in a graph. Moreover, there can be 1G elements that represent so-called \emph{property statements} which resemble the notion of properties in LPGs or of RDF triples with literal objects. Other 1G elements of a 1G dataset may represent so-called \emph{relationship statements} which resemble edges in LPGs or RDF triples that have a resource as object. Every statement of any of these two kinds is associated with a 3-tuple consisting of a 1G element that is the \emph{subject} of the statement, a 1G element that is the \emph{predicate}, and a 1G element or value as \emph{object}.
Finally, every 1G dataset consists of graphs, which
are containers of statements. An important feature of the model is that statements may consist of 1G elements representing other statements, which may be used to capture statement-level metadata and even meta-edges (e.g., edges between edges), similar to the RDF-star extension of RDF~\cite{RDFStarCGReport}.

We emphasize that the abstract 1G model maps naturally to a 4-column relational representation in which the fourth column contains so-called \emph{statement identifiers} (\emph{SIDs}) for the statements captured via the first three columns. Hereafter, these first three columns are named S, P, and O, respectively, as they contain the subject, the predicate, and the object of each statement, and the fourth column~(with the SIDs) is named I. On the logical level, the domains of these columns can be RDF terms (IRIs, blank nodes, literals)~\cite{RDF11Concepts}, for which the graph-scoped notion of identity of elements in LPGs needs to be harmonized with the globally-scoped notion of Internationalized Resource Identifiers (IRIs)~\cite{RFC3987:IRIs} in RDF. To this end, we apply the notion of a {\em baseIRI} associated with each graph dataset, as discussed below. Note that multiple SPOI tuples with the same S, P , and O values are permitted by the 1G model; this supports the LPG concept of multiple edges having the same source, link type, and target. If a query engine that runs in SPARQL mode encounters such data, it must project out the distinct SPO values to reduce the data model to RDF triples.\

%% file: rel.tex
\begin{table*}
  \caption{Relation descriptions, S and O column types, and statement identifier classification (SIDs)}
\vspace{-5pt} 
  \label{table:relations}
  \begin{tabular}{lrl|l|l}
  \toprule
  \multicolumn{3}{l|}{Relations} & S $\rightarrow$ O edge type & I (SID)\\
  \midrule
   Topology:&\emph{edges}            &\RELE   & IRI/BNode $\rightarrow$ IRI/BNode & {\sc Sid\_E}\\
   &\emph{left meta edges}  &\RELLME & SID $\rightarrow$ IRI/BNode & {\sc Sid\_Lme} \\
   &\emph{meta edges}       &\RELME  & SID $\rightarrow$ SID & {\sc Sid\_Me} \\
   &\emph{right meta edges} &\RELRME & IRI/BNode $\rightarrow$ SID & {\sc Sid\_Rme} \\
  \midrule
  Edge Values:&\emph{edge properties}  &\RELEP   & {\sc Sid\_E, Sid\_Ep} $\rightarrow$ IRI/BNode, Literal & {\sc Sid\_Ep}\\
   &\emph{left meta edge properties}  &\RELLMEP &{\sc Sid\_Lme, Sid\_Lmep} $\rightarrow$ IRI/BNode, Literal & {\sc Sid\_Lmep} \\
   &\emph{meta edge properties}       &\RELMEP  & {\sc Sid\_Me, Sid\_Mep} $\rightarrow$ IRI/BNode, Literal & {\sc Sid\_Mep} \\
   &\emph{right meta edge properties} &\RELRMEP & {\sc Sid\_Rme, Sid\_Rmep} $\rightarrow$ IRI/BNode, Literal & {\sc Sid\_Rmep} \\
  \midrule
  Vertex Values:& \emph{vertex properties}            &\RELVP   & IRI/BNode $\rightarrow$ IRI/BNode,Literal& {\sc Sid\_Vp}\\
   &\emph{meta vertex properties}  &\RELMVP & {\sc Sid\_Vp, Sid\_Mvp} $\rightarrow$ IRI/BNode, Literal & {\sc Sid\_Mvp} \\
   &\emph{vertex vector search}  &\RELVSS & IRI/BNode $\rightarrow$ vector & {\sc Sid\_Vss} \\
  \midrule
  Dictionaries:&\emph{vertices}            &\RELVERT & \multicolumn{2}{l}{IRI/BNode mapped to \gui}\\
   &\emph{literals}            &\RELLIT  & \multicolumn{2}{l}{String mapped to \gui}\\
  \bottomrule
  \end{tabular}
\end{table*}

The naive way to define a physical data model for 1G would be a simple 4-column relation, where each of the SPOI
columns are represented in lexical form. Such representation will not be memory efficient as it would necessitate
replicating data in every column. \Peight is designed to be an in-memory storage engine for both OLTP and
OLAP workloads. Thus, data must be stored in a compact way, data locality maximized, and enable
parallelism for higher throughput and scale-out computation. To meet these requirements, we use dictionaries to encode
literals (lexical or binary data) and IRIs. A dictionary encoding maps a IRI or a literal to a 64-bit
global unique identifiers (\gui). We also separate literals and IRIs into two different
dictionaries. This allows for faster localized access of IRIs, since they are separately from literals that can be
very large strings or blobs. Similarly, we separate 1G elements, such as property statements and topology statements, in
different relations to further improve low-latency access via more compact and local relations. We also partition the
relations to enable parallel access, while making sure that we co-locate topology and property relations of the same set
of vertices to facilitate joins between them.

In total, \Peight's storage model is based on 13 relations.
Two \emph{dictionary relations} which provide a mapping between lexical forms and internal 64-bit Global Unique
Identifiers ({\em GUI}s). The two dictionary relations are \RELVERT and \RELLIT, for vertex identifiers and string literals, respectively;
four \emph{topology relations}. Simple edges are modeled in \RELE, and edges which connect meta-information in another family of relations are modeled separately in the relations \RELME, \RELLME, and \RELRME;
six \emph{property value relations}: one to model vertex property sets (\RELVP) and one for vertex
meta-properties (\RELMVP), then one more for edge property sets (\RELEP), and finally three for edge meta-properties
(\RELLMEP, \RELMEP, \RELRMEP);
a \emph{vector relation}, \RELVSS, which maps vertex identifiers to vectors and maintains an Hierarchical Navigable Small World
(HNSW)~\cite{DBLP:journals/corr/MalkovY16} secondary index over those vectors for efficient approximate nearest neighbor search.

Table~\ref{table:relations} presents the different type of relations (topology, values, meta properties).
The resulting system of 13 distinct relations is grouped into logical containers. These containers provide the ability to
{\em co-locate} certain data for efficient access. For example, edges and edge properties are located in conformal
containers to provide efficient access as are vertex properties and meta-properties. Table~\ref{table:containers}
shows how relations are grouped in the same containers, and how they share {\em join indexes} between common SIDs.

\begin{table}[h]
 \caption{Containers of co-located relations and join indexes}
\vspace{-5pt} 
  \label{table:containers}
  \begin{tabular}{lr}
  \toprule
  Container & Join Index\\
  \midrule
   {\sc C\_E}=(\RELE, \RELEP) & \RELE.I $\bowtie$ \RELEP.S\\
   {\sc C\_Me}=(\RELME, \RELMEP) & \RELME.I $\bowtie$ \RELMEP.S\\
   {\sc C\_Lme}=(\RELLME, \RELLMEP) & \RELLME.I $\bowtie$ \RELLMEP.S\\
   {\sc C\_Rme}=(\RELRME, \RELRMEP) & \RELRME.I $\bowtie$ \RELRMEP.S\\
   {\sc C\_V}=(\RELVP, \RELMVP) & \RELVP.I $\bowtie$ \RELMVP.S\\
  \bottomrule
  \end{tabular}
\end{table}

For capturing a basic LPG, the following five relations are sufficient: \RELVERT, \RELLIT, \RELVP, \RELE, \RELEP.  The 1G model
extends that by allowing {\em meta-edges} between edges, {\em meta-properties} both for vertex properties and for edge
properties, etc. SPOI tuples for such meta-edges and meta-properties use the I value (i.e., the SID) of some other SPOI
tuple in either the S or the O position. For efficient storage and access, we divide the edge-related meta information into
three subgroups depending whether the S and/or O are SIDs. Hence, 1G adds seven more relations (Table~\ref{table:relations}), three topology meta relations:
\RELRME (S is a simple vertex, O is a SID), \RELLME (S is a SID, O is a simple vertex), and \RELME (both S nor O are SIDs),
three corresponding property value relations (\RELRMEP, \RELLMEP, \RELMEP), and one more property value relation for the
{\em meta-vertex properties} (\RELMVP). SPARQL named graphs (a subgraph container concept) are handled as meta-properties
annotating edges and property values, as are user defined edge identifiers.

In order to share the same system of vertex identifiers across LPG and RDF data within the (relational) 1G representation,
we {\em logically} prefix each LPG vertex identifier with the {\em baseIRI} of each 1G graph. For example, if a customer
has a graph \texttt{arn://xyz}, then the vertex identifier "1" is logically interpreted as \texttt{"arn://xyz/1"}.  
For the common case where identifiers are local IDs -- as it always is the case for LPG data, which does not exhibit
a built-in notion of global identifiers (as opposed to using the RDF aspects of 1G to refer to an IRI) -- that prefix is
compressed into a single bit in the \gui created for the vertex identifier. 
All storage operations are performed on \gui and the distinction between LPG and RDF identifiers is invisible to those
operations.   This approach not only aligns identifiers from these two models, but it also make it possible to merge or
federate LPG data by relying on the semantics of the RDF data model.

\begin{figure}[ht]
  \centering
  \includegraphics[width=\columnwidth]{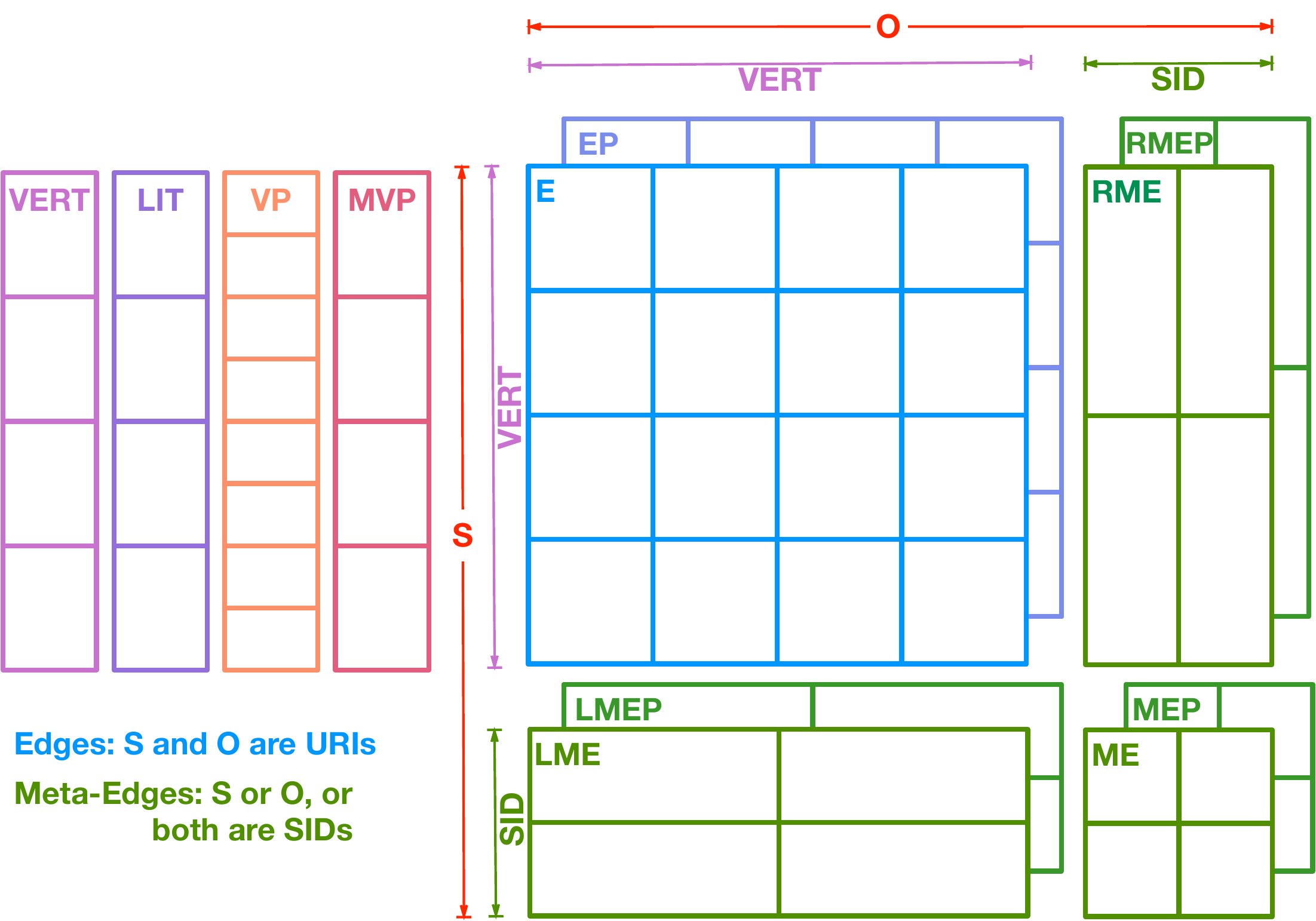}
  \vspace{-8pt}
  \caption{1D and 2D partitioning of \Peight relations.}
  \Description{The 2D partitioning of the relations in \Peight. columns are subjects rows are objects and the 1d value partitions are co-located}
   \label{fig:p8relations}
\vspace{-7pt}
\end{figure}

Having separate relations enables different partitioning schemes for them, as illustrated in Figure~\ref{fig:p8relations}. 
For example, \RELE is partitioned in 2D to optimize multi-level traversals, whereas \RELVP is partitioning in 1D to optimize
locality for fetching vertex properties.

One of the difficult design decisions was how to handle edge weights. Efficient access to edge weights is critical for the
performance of many graph algorithms, such as SSSP. However, 
both the LPG and RDF data models are schema flexible or schema never. In our design, we decompose edges and edge weights into
two different  relations (\RELE and \RELEP) which are conformally partitioned to ensure locality in scale-out architectures. We
also define a join index to make it efficient to lookup the edge weights for an edge. Further, we surface the concept of \RELE
$\bowtie$ \RELEP directly into the data access patterns, making it possible to push down optimizations into low level data scan operations. This design allows to properly model any customer data, including odd cases where an edge property is not always
present, is present with the wrong data type, is present multiple times, is associated with meta-properties about that edge property, etc. To optimize performance for whole graph analytics, when it will be amortized, we build optimized data structures,
such as CSR, which precomputes the edge and the edge weight for maximum performance.


In \Peight, each partitioned relation consist of SPOI tuples (i.e., property statements and edge statements) that are stored in
allocated memory heaps. The partitioned relations are organized as a set of 2MB huge pages using a PAX~\cite{Ailamaki01} format
to support efficient scans. The tuples on each such page are managed by a lock-free free-list for that page. An {\sc Undo}
list is maintained per {\em zone} (a consecutive group of 1024 tuples), extending the concepts presented by Neumann et 
al.~\cite{neumann2015fast}. Secondary indices are also maintained over the partitioned relations using virtual arrays and
lock-free adjacency lists. In addition, an {\em I-index} is defined that provides a lock-free factory for unique SIDs and a
lookup of SPOI tuples by I.

\begin{figure}[ht]
    \centering
    \includegraphics[width=0.8\columnwidth]{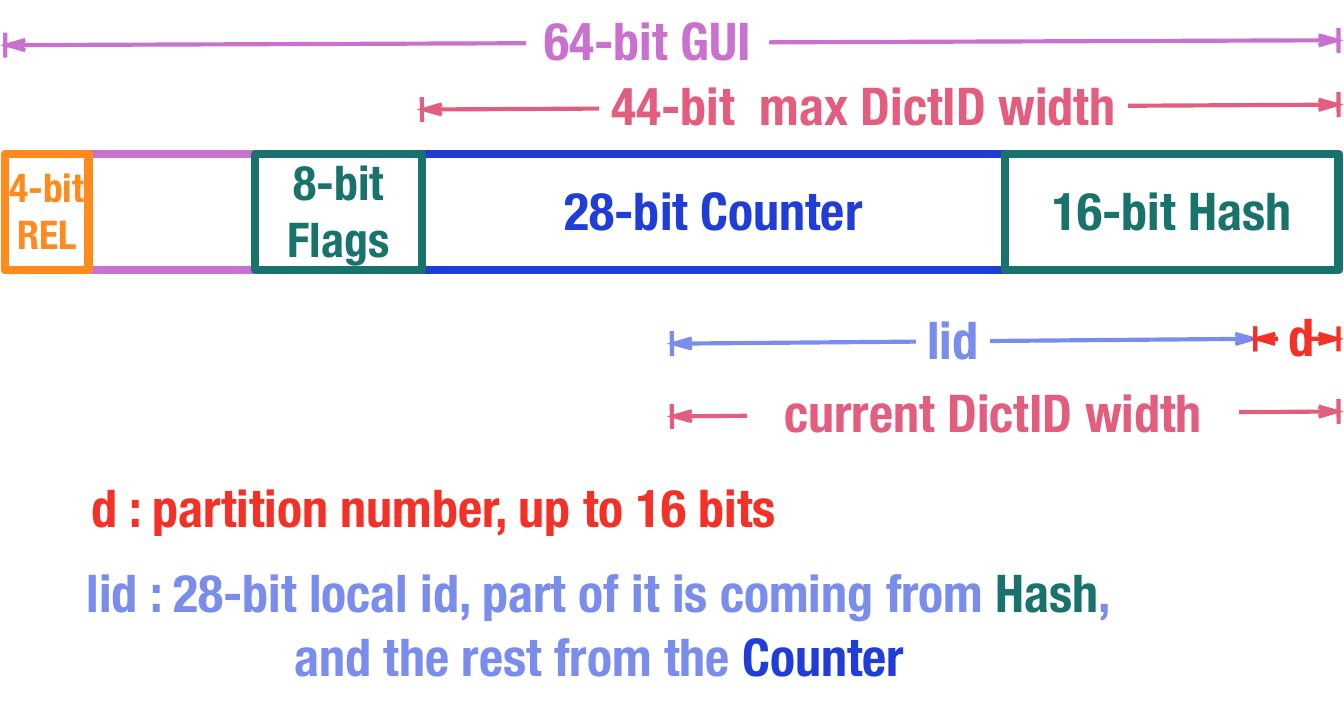}
    \vspace{-8pt}
    \caption{Dictionary encoding to \gui.}
    \vspace{-8pt}
    \label{fig:gui-creation}
    \Description{Dictionary encoding to gui.}
\vspace{-9pt}
\end{figure}

To improve the in-memory data density, all data is stored compressed using 64-bit {\em Global Unique Identifiers} (\gui) and,
whenever possible, using only parts of the \gui. We distinguish two kinds of \gui identifiers: \emph{dictionary \gui}, as are
assigned in the two dictionary relations (\RELVERT, \RELLIT), and statement \gui that represent the tuple 
identifiers \emph{SIDs}. 
Within each \gui, we use four bits to encode which of the 13 relations the \gui belongs to; for a dictionary \gui, that is the
dictionary relation in which the \gui is assigned, and for a \gui representing a SID, it is the relation containing the SPOI
tuple in which that SID is the I value. The rest of the encoding depends on the kind of \gui. Dictionary \gui consist of a 
44-bit identifier that is unique within the corresponding dictionary and is generated by Murmur hashing (16 bits in LSB positions) combined with a 28-bit collision counter, as illustrated in Figure~\ref{fig:gui-creation}.  
In contrast, \gui for SIDs consist of a 28-bit partition-local identifier (obtained from the partition-local I-index), plus 32 bits to identify the partition.\footnote{This means there can be a maximum of $2^{32}=4$B partitions and each partition can contain at most $2^{28} = 256$M 1G elements. These limits come from the design choices based on our desire to limit the system recovery time, as well as to improve parallelism.} 

Most relations use a 1D partitioning system where the low bits of the S determine the partition in which SPOI tuples for that S
will be located (see Figure~\ref{fig:gui-creation}, where $d$ bits are used for partition number) and all SPOI tuples for the 
same S are in the same partition. Furthermote, 2D partitioning is applied to the topology.  
This has several advantages (facilitating parallelism, load balancing, increasing the maximum degree of a vertex, minimizing contention in algorithm kernels), but also some trade-offs (adjacency lists of vertices are partitioned).  
The 2D partitioning system creates a grid of partitions where the rows of the grid are indexed by S and the columns of the grid
are indexed by O. Thus, if only S is bound, then the adjacency list for that S is distributed across a row of the 2D grid.
Likewise, if only O is bound, then the adjacency list for the O is distributed across a column of the grid.  
If both S and O are known, then the edges for that (S,O) are guaranteed to appear in a single partition.  
The 2D partitioning makes it possible to elide the low bits of each S and O value in the topology since those bits correspond to
the row or column of the grid, thus we store S and O in 28 bits within \RELE.  
Using similar logic, we store S in 28 bits within the partitioned of \RELVP and the other non-meta relations (for meta relations
we store the 64-bit S and O values). Also in all 1G relations have a partition local encoding of predicates, and hence the
predicate is also stored in 4-bytes. The property value partitions use 9 bytes to store the O, including an 8-bit flags field,
indicating type, and either inline primitive types up to 64-bits or the GUI of larger values. In total, topology (\RELE) tuples
are only 16-byte wide and property value tuples (\RELEP, \RELVP, etc.) are 21-byte wide. 

%% file: storage.tex
\Peight\ supports a variant of "fast MVCC" ~\cite{neumann2015fast}, but with modifications to support
a graph data model. Each zone (of 1024) rows is associated with a version vector that marks tuples within that zone
having {\sc UNDO} information. The {\sc UNDO} chain contains the pre-image
version of the tuple. Due to their narrow width and the nature of the 1G data model, tuples are
only inserted or deleted, not modified in place. Therefore, {\sc UNDO} chains are at most 1- or 2- 
deep (insert only, delete only, or insert + delete). SPOI tuples are bit-marked if they
are on the free list (a lock-free data structure which is intrusively linked for tuples on 
the same page) and if they are associated with multi-version information. Because the S and O
values are based on a murmur hash and collision counter, neighboring values are not "close" in
the application semantics. Therefore, rather than track {\em VersionedPositions}, we steer 
data access (scans) around versioned tuples using other mechanisms. Older versions are expunged
once the Oldest Active Transaction (OAT) can no longer read a tuple.

All transactions are tracked in a transaction table. They are assigned a handle based on their position in the transaction table. A transaction is assigned a read timestamp and each successful read-write transaction also gets a commit timestamp. The timestamps are linearly ordered values stored in the transaction table, and commit timestamp is incremental and unique. The durable logical log records are written incrementally after in-memory writes complete, and reference the transaction handle, which can be safely recycled after server restarts. A transaction abort and commit is recorded with the transaction handle as well, an the commit record carries the commit timestamp as well. Incomplete transactions at the tail of the log are discovered on restart and explicitly aborted.

Undo logs are assigned to a chain of memory aligned blocks, and each block is used exclusively for a single transaction. The transaction handle is stored once in the header of the blocks. When a SPOI tuple is read with multi-version bit set, we get the transaction handle in the header of the block and from there go to the transaction table to get the state of the transaction (in progress, committed, aborting, terminating, etc.) and commit timestamp for a committed transaction. By checking the state of transaction and commit and read timestamp comparisons, we decide if a tuple is visible to a transaction.

We use another index to track read timestamps of actively running transactions and commit timestamps of committed transactions that need to be purged. The read timestamps of active transactions are reference counted. Whenever the reference count of a read timestamp becomes zero, we check whether we can advance the OAT.  If the OAT is advanced, we purge any committed transactions with commit timestamp lower than the new OAT, including expunging any tuples versions which are no longer visible to any active transaction.

%
%

One of the early design decisions was to use a durable {\em logical} log.  This has proven to be 
a tremendous benefit in developing a new storage engine because it completely separates
the manner in which the data are organized for durability from the manner in which the data are 
organized for compute.  This separation of concerns has made it possible for us to re-invent the
in-memory layout and indexing methods on an ongoing basis, improving load rates, reducing memory
stalls, reducing memory consumption, improving handling of hypersparse graphs, etc. without any
concerns about breaking durable storage.

The durable log is fronted by a chain of lock-free box cars which write onto a partitioned log.
Each partition of the log can accept up to a fixed write rate.  By partitioning the log, we are
able scale the write throughput to the network bandwidth.  We accept higher tail latency associated
with a partitioned log because there is little locality in graph updates and, unlike relational
workloads, nearly every query or update will touch multiple partitions.


Restore is from the most recent set of S3 checkpoints, followed by the apply of the log.  Like writes,
restore is able to saturate the network bandwidth.  Recovery and log apply are parallel and recovery 
time is bounded by the work each thread needs to do to restore its partitions.

%% file: index.tex
We support two methods of scan operations to retrieve data from storage. One is based on performing a complete
scan over the relations heap, and the other through the use of indexes that map vertices to their adjacency list
of edges. The choice between these two access methods -- relation scan vs index scan -- is based on a combination
of factors: the type of relation (topology or values), the type of data (resources, strings, or inlined 
numericals), and the selectivity of the input filter. For example, if the vertices that are requested to be
scanned have a high degree or if the query has a very low selectivity rate then we will opt for scanning the
relation from the heap instead of going through an index scan. This is because we expect to read most of the
data from the heap and produce many results, thus sequential scans will outperform index probes.
On the other hand, if we are given a point query, or a small frontier of a set of vertices, then an index scan
over the adjacency list index will be preferred. 

The input interface of all scan operations, including the filters, frontiers for joins, and projections, is
based on the specialized language \APL which we will detail in the next section. It is also part of the \APL 
implementation the mechanism in which we decide the use of full relation scan versus index scans, or even a 
combination of these two.

\paragraph{Relation Scans}
\label{subsec:relationscans}

As described previously, each relation is divided into partitions and each partition is stored in the memory heap.
Each partitioned relation is divided into pages, where each page is an allocated 2 Megabyte huge page
in memory. Consequently, each partitioned
relation can have up to 2048 pages, and each page can have up to 131K SPOI tuples, totaling $2^{28}=268$ million
tuples per partition (for \RELE with 16-byte wide tuples). Each page is then organized into a 
PAX~\cite{Ailamaki01} layout. The PAX blocking of a page, i.e. the way that columns are grouped together, depends
on the type of relation, and also on the priority of the application (OLTP vs OLAP). We can block together all
the columns for  faster point queries, or break up columns into groups for faster sequential scans on a single
column. For example, in the case of topology, we group together in a 16-byte value all the SPOI columns to create 
an OLTP bias, while in the case of a value relations,
we group together SPO, while we store separately the I column and the F column (F column is used to mark the type
of the inlined value in O and is explained below).

Relation scans are sequential scans that follow the memory data layout (blocking) of the columns in order to
achieve optimal memory access. Such scans maximize memory bandwidth reads because they facilitate pre-fetching
and thus reduce cache and TLB misses. The relation column scan operation works by scanning each page of the 
partitioned relation following the blocking structure of the columns (i.e, the PAX layout) and by materializing
the results for the current segment before proceeding to the next segment. Therefore, the relation scan algorithm
can be viewed as a vectorized execution, where each vector is of the size of the PAX page.

We outline the relation scan algorithm for the general case, albeit small changes in the order of the columns
scanned inside each page, subject to the different blocking layouts. The relation scan algorithm starts by
looping over the pages in the partitioned relation. For each page, first the most selective column is scanned and
a bit-vector is produced, such that the corresponding bits are set if the value in the scanned column matches the
input filter or frontier. For the next column in the same page, we iterate over the bit-vector instead of the
base data, and if the bit in position $i$ is set, then we check the value in \texttt{col}[$i$] if it matches
the input filter and if so we continue, otherwise we set the bit $i$ of the bit-vector to 0 and ignore that
tuple. At the end, the algorithm has produced a bit-vector with bits set only for the SPOI tuples that
satisfy the input filters. We use that bit-vector to materialize the results and continue to the next segment.
In order for the relation scans to perform optimal, we avoid function calls and if-statement branching. For
these reasons, our code relies heavily on C++ templates. The templates implement the access patterns as defined and
enumerated by the \APL interface. We also plan to extend our scan code with JIT compilation techniques to 
incorporate complex math computations, user defined functions, and aggregations with group by operations.

For the value relations where we are interested in the type of the literal values, which can be either strings or 
numerical, we introduce one more column, the {\sc Flag} array that serves to identify the type of the O column.
The {\sc Flag} array has one 8-bit value per SPOI, effectively transforming SPOI to SPFOI. The purpose is to 
quickly identify the type of O during scan operations and thus skip over data that do not match the query filters,
or to quickly decide if a type can be promoted to another type, for example a short to integer.
The {\sc Flag} array divides literals in O column not only according to their base type (e.g., integer, double, 
IRI, etc.) but also into larger groups of types, such as numerical vs non-numericals,
composite vs non-composite, and also subgroups such as signed and unsigned, etc.
\paragraph{Index Scans}
\label{subsec:indexscans}
Besides relation scans, which are mostly useful for read-optimized queries, we also support point queries
useful for transactional workloads. For this reason we have defined three indexes over columns S, O, and I. 
For the S and O column the indexes are high concurrency hashmaps of vertices mapping to adjacency lists where lock free
updates are possible. The index on I column is a join index that joins the co-located relations as described in
Section~\ref{sec:rel}. For column S, the hash index maps a vertex identifier in all partitions, while for column O, the
hash index is a vertex identifier for both the topology and wide topology partitions. For values relations, O is either
a dictionary identifier for the case of string literals, or a resource (IRI). Notice that we do not include numerical values
in the O index although in future work we plan to implement a specialized order preserving index.

The adjacency lists provides a lock free data structure which maintains a direct list of all tuples in the
partition having that dictionary identifier (\gui) in the corresponding position.  The adjacency list is a contiguous
allocation with an exponential growth policy. When an adjacency list becomes full, it is compacted, deleted entries are
purged (a delete marker is tracked), the size of the new adjacency list is determined, and the entries are copied over to
the new adjacency list and then sorted again for physical locality. This sort order minimizes the likelihood that scanning an
adjacency list will induce TLB misses. The index of the last sorted position in the adjacency list is tracked and scan
primitives can use this to optimize their behavior. The new adjacency list is installed into the hashmap index using
an update/insert operation, which is based on a compare-and-swap instructions. In a data race, one thread wins and the other 
thread(s) discard their uninstalled version of the new adjacency list.


%% file: apl.tex
\input{aplTable}
Amazon Neptune is built with a collection of different components that can be stacked to build user applications.
In order for \Peight to be part of this software stack we developed a data retrieval interface for that is
based on a declarative but frugal language. This language---called the \emph{\Peight access pattern
language} (\APL)---can be used to describe data retrieval requests over the logical SPOI relation used as
the basis for representing 1G data in \Peight~(i.e., the language is agnostic to the physical data model).
\APL hides the internal data structures, index availability, and execution plans from the caller component,
thus making \Peight easy to integrate and use within Neptune. Given a data retrieval
request, an \APL processor creates a concrete data access plan for the request, runs this plan to fetch the
requested data, and writes the results to a memory location specified as part of the request. To determine
such a plan the \APL processor employs a (rule and cost based) optimizer for making decisions about suitable
access paths, and about usage of secondary indices such as bloom filters, imprints~\cite{Sidirourgos13}, 
etc. The remainder of this section provides an informal overview of \APL and, thereafter, briefly describes
the \APL~processor.

\subsection{\Peight Access Pattern Language (\APL)} \label{ssec:apl:language}

\APL expressions are meant to be used as arguments for data retrieval operations---hereafter, called \emph{\APL requests}---that fetch data from the \Peight relations. Additional arguments for such a request are pointers to data structures that either contain constants that are meant to be bound in a particular position of the specified access pattern (effectively supporting filters and frontiers) or represent the data structures into which the fetched data has to be retrieved.
Our design goals for \APL are: 
\begin{itemize}
	\item
		\APL
	should be able to \emph{capture the range of access patterns} found across different query languages, while offering an extensible, query-language-agnostic
	abstraction layer.

	\item
	\APL should be declarative, abstracting both from the separation of SPOI tuples into multiple relations on the physical level and the corresponding partitioning scheme.

	\item
	\APL expressions can be \emph{parsed efficiently} because parsing them is part of the critical path during query execution.

	\item
	\APL expressions can be \emph{very concise} and do not need to be easily readable for humans, because the language is designed to be generated by the query engine and interpreted by the storage layer.
\end{itemize}

\paragraph{Overall Structure of a \APL Expression.}
Table~\ref{tab:APLExamples} lists a few examples of \APL expressions to provide an overview of the main features and the syntax of the language. Related to the four main attributes of the SPOI relations, \APL expressions consist of four main components: the subject component~(e.g., "\verb|S|" in example expression~\#1 in Table~\ref{tab:APLExamples}), the predicate component~("\verb|P|" in expression~\#1), the object component~("\verb|_^|" in~\#1), and the SID component~("\verb|_|" in~\#1). Within these components we use different annotation characters to define, e.g., if an attribute is to be projected or not in the retrieved data (the character to request projection is: \verb|^|), if it is unbound (character: \verb|_|) or bounded to a given set of constants (for filtering) or to a frontier (for joins against a known set of values%
)%
	, if a type flag is available (for exact type matching),
~etc.
For \APL expressions that define access patterns to retrieve data from multiple columns of the SPOI relations (e.g., example expression~\#2), the lists of values in the corresponding output data structures will be aligned with one another. That is, the $i$-th value in each of these lists will be obtained from the same~SPOI~tuple.

\paragraph{Filtering based on Values.}
An access pattern with a set of fixed values for a particular column can be specified by using the corresponding column name~(e.g., "\verb|S|" and "\verb|P|" in example expression~\#1). 
Another option is to use the sub-expression "\verb|[]|" (as in expression~\#2), which indicates that a list of possible values for the corresponding SPOI column is passed within the \APL request and, then, considered in the access path as a frontier. 
The difference between these two options (e.g., "\verb|S|" versus "\verb|_[]|" in examples \#1 and \#2) is subtle: the first one refers to constant value(s) while the second to a frontier. Almost always, a frontier has to be evaluated with a (hash-)join, whereas constants can be handled either with joining or with sorting and probing, depending on the cardinality, statistics, and placement. Exactly these choices are meant to be hidden from the caller and left to the \APL processor to decide. Notice that it is also possible to write "\verb|S[]|" in order to indicate that both a set of constant values for filtering and a frontier are given.

\paragraph{Type Restrictions.}
For some cases, the query engine built on top of \Peight may infer that the S or the O value of matching SPOI tuples can be restricted to a particular kind of values. For instance, if the query to be executed by the query engine retrieves
	edge that have
edge properties, then the access pattern for this part of the query can be restricted to SPOI tuples that have a SID as their S value and a literal as their O value. Example
\#3 illustrates this case.
The \APL planner in \Peight uses such type restrictions to prune irrelevant access paths%
	~(e.g., for example~\#3, only the relations \RELEP, \RELMEP, \RELLMEP, and \RELRMEP are relevant).

\paragraph{Additional Sub-Pattern for the SID Component.}
In place of the SID component, a \APL expression may contain another \APL expression, as illustrated by example~\#4. Such types of combined \APL expressions describe access patterns that resemble an equi-join with a predefined join condition (namely, the I value of the left-input tuples must be equivalent to the S value of the right-input tuples). The letter "\verb|x|" between the two sub-expressions indicates the use of an inner join, whereas the letter "\verb|k|" would indicate a left outer join.
The use case of such a combination of \APL expressions is to capture typical patterns of joint data retrieval in a single \APL request; e.g., retrieving edges together with their properties, or retrieving properties together with meta-properties about them.
While a query engine on top of \Peight may capture such joins also in its query plans, supporting them within a single \APL request can result in more efficient query plans because \Peight may, e.g., exploit partition co-location for such joins.

\subsection{\APL Planner} \label{ssec:apl:processor}

\APL defines a finite number of different access patterns for retrieving data. However, these access patterns can entail different execution plans. The most common approach
is to use statistics to predict the selectivity of an access pattern and thus choose index or relation scans. However, we can also employ heuristics in order to decide the
order in which we scan a column. Such heuristics have been studied in~\cite{Tsialiamanis12} and are also applicable in our design. 

The \APL planner is responsible for firstly determining which partitions need to be scanned subject to the existence of bounds in either S or O or both. 
The planner will then determine if a relation or index scan will be used per partition based on the cardinality of the set of vertices that can be found in each partition. Finally, the
order of column access in the case of relation scans, or the usage of S or O index in the
case of index scan is decided. In some cases, the planner may also choose to use both scan
options and combine the intermediate results to reduce the data accessed or to first
evaluate Property predicates or inline values and then the adjacency lists of a set of vertices.

In addition, the \APL planner will inject a {\em gather} operation if needed, in order to
collect intermediate results from different partitions and compute a sum, or group together
edges with a group by parameter different from the indexed vertex. Finally, the planner will
stop the execution of a scan if a \textsc{Limit} or \textsc{Sample/Count} operation is
present in the query.

%% file: aplTable.tex
\begin{table*}[t]
	\caption{Examples of \APL expressions, including the additional arguments expected when using them, and their meaning.}
 \vspace{-5pt}
	\label{tab:APLExamples}
	\begin{tabular}{ccp{65mm}p{75mm}}
		\toprule
		\# & Expression & Meaning & Additional Arguments\\
		\midrule
		1
		& \verb|SP_^_|
		& retrieve the O value of every SPOI tuple that has a given S value and a given P value
		& i)~value for S, ii)~value for P, iii)~pointer to data structure to which the O values of the selected tuples are written
		\\ \midrule
		2
		& \verb|_[]P_^_^|
		& retrieve the O and the I value of all SPOI tuples that have both a specific P value and an S value that is one of the elements in a list of such values
		& i)~value for P, ii)~pointer to data structure with frontier for S; iii--iv)~pointers to two data structures to which the O and the I values of the selected tuples, respectively are written
		\\ \midrule
		3
		& \verb|s_^_l__|
		& retrieve all SIDs that are S values of SPOI tuples with a literal in the O position
		& i)~pointer to data structure to which the S values of the selected tuples are written
		\\ \midrule
		4
		& \verb|SP_^x__^_^_|
		& join every SPOI tuple $(s, p, o, i)$ with every SPOI tuple $(s'\!, p'\!, o'\!, i')$ such that $i=s'\!$, and retrieve $o$, $p'\!$, and $o'$ of every result tuple 
		& i--iii)~pointers to three data structures to which the three values of every result tuple are written, respectively
		\\
		\bottomrule
	\end{tabular}
\end{table*}

%% file: algos.tex
Neptune Analytics provides a managed service for customers to run top-k (BFS, EgoNet, etc.), whole graph single pass algorithms (e.g., BFS, SSSP, WCC, etc.) and whole graph analytics (PageRank, Louvain, etc.) over live and dynamic graph data and achieve performance that is comparable with the state-of-the-art HPC graph analytics implementations. 

Developing high-performance parallel graph algorithms
is a challenging task that requires addressing several well-known hardware and
software-related challenges~\cite{Lumsdaine07-PPL}.
Graph algorithms have random data access patterns, and since the ratio of computation to data access very low, they are memory-bound.
Furthermore, graph algorithms are data-driven and most of the real-life graph datasets have very skewed distributions.
This causes a significant load imbalance challenge during algorithm execution. 
Literature shows that use of 2D partitioning can provide a sweet-spot to compromise some of these challenges~\cite{Yasar22-ARXIV}, if not all.
Hence, as described in Section~\ref{sec:rel}, in \Peight we adopted 2D partitioning of the topology relations. 
In each partition, SPOIs are stored in our heaps can be viewed as Coordinate List (COO) format, which is one of the common formats in graph analytics, since each edge's two "coordinates", that is, S(ubject) and O(bject) are stored explicitly.
Even though this could be an effective storage for some of the graph algorithms, especially on GPUs for edge-centric execution,
most of the CPU-based HPC graph analytics systems (e.g.,~\cite{Shun13-PPOPP}) and individual algorithm implementations (e.g.,~\cite{Catalyurek12-ParCo,Yoo05}), including recent baseline benchmark implementations (such as GAPBS~\cite{beamer15-gap}), work on static Compressed Sparse Row (CSR) or Compressed Sparse Column (CSC) representations of the graph, which are one of the most compact and efficient structure to access the adjacency list of the vertices.

The interplay of data layout, dynamic data, and algorithm characteristics and their effect on performance is important to consider.  
For example, top-k analytics and algorithms that will do only a single pass over the graph run against the most recently committed state of the graph.  
This enables a wide range of applications where the recent information can be critical, such as detecting fraudulent transactions.  
When the cost of building static data structures can be amortized, such as in Louvain or multiple concurrent top-k queries, we build partition-local CSR or CSC data structures, for algorithms that will do multiple passes over the data.  
These ephemeral CSR data structures are constructed from the current committed state of the graph as live data as a {\em static view} containing all data that should be visible to algorithm. 

Graph algorithms are implemented using {\em graph kernels} that abstract how graph data is stored and accessed, and leaving the algorithm developer free to focus on the algorithm's input and output, ephemeral state, and the operations that needs to be carried out during the algorithm. 
This is done by implementing few algorithm specific call-back functions: such as initialization of ephemeral data; what needs to be done when a vertex or an edge is visited, etc. 

The data abstraction and foundations of main loop constructs for iterative algorithms are implemented in two 
core framework functions:
(1)~{\em OneHopExpansion} for a given part and a {\em frontier}, i.e., a set of input vertices, visits the neighbors of the vertices in the frontier, and calls the graph kernel's appropriate call-back functions;
(2)~{\em ItarativeAlgorithm} starting with a given frontier, coordinates execution strategies (sequential, 1D parallel, 2D parallel) of OneHopExpansion for all relevant parts, collects the output frontiers, and continues to iterate level-by-level converting current output frontiers to the next level's input frontier, until the stopping condition (another graph kernel call-back function) is satisfied.
By taking into account input from Iterative Algorithm, and the 
properties of input frontier and data in the partition, OneHopExpansion
chooses a traversal mode that is either index-driven, or scan-driven.
Both scan and index-driven execution are carefully optimized to read 
only the data needed for the execution by taking advantage of PAX 
columnar data storage, as well zone-based version vectors, to ensure a 
transactionally consistent view of the data based.

The traversal mode can be further optimized for single pass, path finding algorithms, such as BFS, by dynamically changing the mode based on the current state of the algorithm, which we call the {\em hybrid} execution mode (which is the default).  Such algorithms generally start with a single {\em source vertex}, hence execution can start with index-driven mode, as the frontier gets larger, can switch to scan-driven execution to achieve higher performance, and in later stages when frontier gets smaller, can revert back to index-driven mode.

One of the key call-back that a graph kernel needs to implement is the visitor call-back.
Three variants with different call parameters exist for optimizing data access: (1)~ target vertex, (2)~source and target vertices, (3)~source and target vertices and edge weight.
Some algorithms, such as BFS (if we are not tracking parents), only requires the target vertices, which will be added to output frontier.
Other kernels may require both source and target vertices, as well as the weight of the edge between them, requiring a fast join between \RELE and \RELEP.

Graph kernels can have 1D or 2D partitioned ephemeral data, conformal with the current topology relation partitions.
For example, BFS level or the PageRank of a vertex are stored in 1D partitioned ephemeral data structures.
Even though topology uses a 2D partitioning system, most of the current algorithms are parallelized over 1D (for S or O), depending on the write/update-pattern of the algorithm, in order to take most efficient sequential data structures to store/aggregate ephemeral values for vertices.

%% file: queries.tex
We launched Neptune Analytics with support for openCypher, a popular declarative graph query language with extensions for invoking graph algorithms.  We then followed up that initial launch with {\em OneGraph} support for RDF, allowing customers to load and query RDF data using openCypher.  The Neptune Analytics stack is explicitly designed to be query language agnostic. This provides us with a path to onboard additional graph query languages like Gremlin, SPARQL, and GQL in the future.

At its core, openCypher provides an ASCII art syntax for matching patterns against the input graph. On top of these basic graph matching facilities, the language offers constructs that map to Relational Algebra like operators such as projections, filters, aggregations, and different join flavors (natural join, left outer join, join exists, etc.)~\cite{marton2017formalising}. Its type system comprises standard primitive types (boolean, numerics, strings), graph specific types (nodes, relationships), as well as composite types such as lists, maps, and paths. Algorithm invocation is supported via the {\tt CALL} clause, which is openCypher's built-in mechanism to invoke stored procedures.   

The following is an example of an openCypher query that computes the Jaccard similarity between {\tt Jane Doe} and {\tt John Doe}, extracts all relationships between them, folds the relationships into a list, and reports back the score as well as the relationship list:
{\footnotesize	
\begin{verbatim}
 PREFIX foaf : <http://xmlns.com/foaf/0.1/>
 MATCH (jane { foaf::firstName : 'Jane', foaf::lastName : 'Doe' })
 MATCH (john { foaf::firstName : 'John', foaf::lastName : 'Doe' })
 CALL neptune.algo.jaccardSimilarity(jane, john, {})
 YIELD score
 OPTIONAL MATCH (jane)-[rel]-(john)
 WITH score, collect(rel) AS relationships
 RETURN score, relationships
\end{verbatim}
}
In addition to providing a look and feel for the openCypher query language, the example aims to demonstrate two Neptune Analytics specific aspects of openCypher usage: (1)~the {\tt PREFIX} keyword introduces a namespace {\tt foaf}\footnote{{\tt foaf} is a commonly used namespace for RDF data that provides shared vocabulary to describe persons and their relationships, see \url{http://xmlns.com/foaf/spec/}.}, which is then used to specify RDF specific node labels ({\tt foaf::Person)} and properties ({\tt foaf::firstName}, {\tt foaf::lastName}); it is a syntactic openCypher extension to ease access of global identifiers (IRIs) originating from RDF data loaded
; (2)~the query illustrates the seamless integration of algorithm invocation via {\tt CALL} clause and traditional querying.

Queries
run through a holistic query processing pipeline. Processing stages include parsing of the input query, the transformation of the parsed AST into a language agnostic logical AST, the optimization of this logical AST, the translation of the optimized AST into a physical plan, and the execution of the physical plan. As a notable special case, for so-called {\it parameterized} openCypher queries -- a protocol-level feature that enables the separation of input parameters from a parameterized query template, akin to SQL prepared statements -- Neptune Analytics leverages a query plan cache that skips parsing, transformation, and optimization in lieu of direct instantiation of pre-computed physical plans, which is crucial to minimize redundant work and accelerate OLTP workloads with high throughput and low latency requirements.

While an in-depth description of the query processing pipeline is beyond the scope of this paper, in the rest of this section we sketch key interfaces between the query layer and the \Peight backend. 

\subsection{Statistics and Cardinality Estimation}

Starting with a discussion of the optimization layer, Neptune Analytics optimizes queries through a sequence of rewrites. Inspired by algebraic equivalences over a graph algebra~\cite{schmidt2010foundations}, these rewrites include general normalizations (e.g., join group flattening), redundant pattern optimization, and algebraic reordering of the logical AST using techniques such as filter pushdown, projection pushdown, and, at its core, join order optimization. Finding a good join order is particularly challenging in graphs. The key challenges are that (C1) from a logical perspective, graph data is fully decomposed (into edges and properties -- as opposed to wide relations encountered in Relational Databases that group related attributes together) and the re-composition of this decomposed data often induces a {\it large number of joins} (with tens of joins per query being the norm rather than the exception); (C2) estimating join cardinalities during the re-composition process demands innovative approaches to statistics that {\it capture correlations in the decomposed relations}; (C3) power nodes introduce {\it skew in the data that causes high variance in result cardinality}. To illustrate the latter by example, the computation of the three-hop neighborhood for a person in a social network -- a sequence of two join expansions from the starting person -- may return very few to millions of results, depending on whether the query starts out from (or, during expansion, encounters) a power node. Another common challenge is that (C4) intermediate cardinalities are often highly {\it sensitive to the traversal direction} (e.g., an edge such as {\tt isPartOf} may expand one-to-one when traversed forward but one-to-many when traversed backwards). To address all these challenges, the optimizer leverages a combination of schema and instance level statistics exposed by the \Peight backend.

{\bf Schema level statistics.} Characteristic sets~\cite{neumann2011characteristic} capture structural patterns in the data -- concretely, aggregated information about nodes that share the exact same sets of properties and (in our implementation: outgoing) edge labels. In addition, they also provide frequency distributions for these structures. For instance, a characteristic set may capture the information that there are 1M vertices in the graph that have exactly one {\tt foaf::firstName} property, one {\tt foaf::lastName} property, and, on average, three outgoing {\tt foaf::knows} relationships. As such, characteristic sets provide concise cardinality estimates for more complex join groups and capture correlations between predicates and edge labels in the input graph (challenges C1, C2). The \Peight backend extends the original notion of characteristic sets proposed in~\cite{neumann2011characteristic} by additionally providing (a)~min and max frequency distributions, which enables the optimizer to assess and quantify the risk of encountering power nodes during traversals (C3) and (b)~HyperLogLog sketches~\cite{flajolet2007hyperloglog} that approximate the number of distinct values for a given property or edge, which helps the optimizer estimate join ratios when estimating join orders that traverse in backwards direction (C4).

{\bf Instance level statistics.} Characteristic sets capture structural information about nodes, properties, and edge labels, but do not provide any {\em instance level} information. Real-world queries, however, often reference specific nodes. In such cases, it is crucial to obtain cardinality estimates that are specific to the given node, which allows the planner to disambiguate power nodes from moderate or low fan-out nodes. To support the extraction of such instance specific estimates, 
the \APL computes cardinality estimations for (primitive, non-join) access paths via sampling -- essentially, approximating the output cardinality of an \APL expression instead of executing it. Complementing the structural information provided by characteristic sets, instance-level cardinality estimates provides additional information to tackle the challenges of skew (C3) and traversal order (C4) for specific starting points provided in the input query.

\subsection{Cost Function and Plan Exploration}

The cardinality estimates that are derived from schema and instance level statistics are used as input to the cost function of the join order optimizer, which drives a Selinger-style~\cite{selinger1979access} dynamic plan exploration algorithm with aggressive cost-based pruning. Within this exploration process, the cost function is used to incrementally estimate the execution time of joins while choosing the best available physical join operator implementation based on cost estimates. 

The physical join operators themselves are configured using \APL expressions, which are (as discussed in Section~\ref{sec:apl}) partitioning scheme agnostic. For access paths that are mapped to relation scans (cf.~Section~\ref{subsec:relationscans}), the execution time is an estimate of the time required to scan (relevant parts of) the relations required to execute the access path; for index scans (cf.~Section~\ref{subsec:indexscans}), we approximate the execution time by summing up the cost of three operations, namely:

{\small
\begin{verbatim}
  (indexLookupTime + adjacencyListScanTime) + projectionTime
\end{verbatim}
}

First, {\tt indexLookupTime} captures the time required to locate, for each incoming value in the input frontier, the adjacency list(s) of the respective value; for the above mentioned case of joining against \RELE, this component depends on the number of \RELE partitions; for instance, looking up a value against the S-index or O-index of an $NxN$ partitioned \RELE relation requires $N$ lookups in total.  Thus, while the cost constant for indexLookupTime is very low, the cost increases logarithmically with the number of \RELE partitions in a row (or column) of the topology relation \footnote{To minimize that cost constant, we replaced the use of concurrent hash maps with direct access arrays backed by virtual memory allocations.  To further reduce cost artifacts and improve space utilization for hyper-sparse graphs, we are in the process of introducing variant of the secondary indexing mechanisms described above where the adjacency list is only partitioned for high degree vertices.}. The second component, {\tt adjacencyListScanTime}, approximates the time required to scan through the adjacency list(s) that were identified through the index lookups and filter out false positives (which may occur, for instance, when the \APL request contains additional constraints on the $P$ position). Third, {\tt projectionTime} captures the time required to copy all matching data into the produced output columns; this component primarily depends on the estimated join output cardinality as well as the number of projected positions.


\subsection{Query Execution}

When it comes to the execution of the physical plan, the \Peight storage and indexing layer is integrated with a purpose-built query execution engine called {\em dataflow execution engine} (DFE), which can be characterized as an in-memory, columnar, vectorized execution engine that supports pipelined query execution through a set of RISC-style operators~\cite{chaudhuri2000rethinking,neumann2008rdf}. DFE execution plans are represented as directed acyclic graphs whose nodes represent operators and the directed edges represent the flow of data across them.  

DFE operators fall into either of two categories: those that (A)~operate directly against the data served by the \Peight backend (i.e., operators for scans and lookups against the base relations, bi-directional resolution of terms against the dictionary, and operators for the specific graph algorithms as discussed in Section~\ref{sec:algos}) vs.~(B)~operators that operate on (intermediate) in-memory results (e.g, selection, filter evaluation, union, or aggregation). The clear separation of operators that access base graph data vs.~those that operate over intermediate solutions results in a clean abstraction layer and makes DFE easily re-usable over different storage backends, by just swapping out the category~(A) operators; in fact, we are using the DFE engine with a different set of data access operators as the default openCypher execution engine for the original Neptune Database product offering that precedes Neptune Analytics (the original platform emphasises OLTP workloads, scales to graphs much larger than memory, hosts three different query languages, but does not support graph algorithms).  Finally, graph algorithm operators support internal parallelism.

The DFE engine can be configured to use either single-threaded execution (implemented via co-routines, using a depth-first execution paradigm) or ~multi-threaded execution (push/push-based, with each thread taking care of scheduling the next action for that thread when it blocks needing input chunks or fills up an out edge buffer). Taken together, these two execution modes capture the spectrum required for an HTAP engine: single-threaded execution is particularly useful for OLTP workloads with high throughput and low latency requirements (no thread coordination and context switching overhead) and pipelined {\tt LIMIT} queries (due to its depth-first execution paradigm that minimizes time to first results), whereas multi-threaded execution is particularly attractive to overlap computation and hide latency when executing compute intensive, OLAP-style queries.

%% file: perf.tex
\paragraph{Scan Rates}
\label{sec:perf-scan}
To quantify the performance of \Peight we compare both index scans and relation scans. We use the 
LDBC Social Network Benchmark (SNB)~\cite{angles2024ldbcsocialnetworkbenchmark} (scale
factor 10). We deployed a Poseidon instance with a $16\times 16$ 2D grid partitioning for the
topology relations. All experiments were run on an Amazon R6i instance. For our queries we
choose three different vertex types with a bound P value and fired \APL scan requests with
different sized frontiers. Each of the three vertex types have a different average degree,
thus varying the number of results per vertex. Specifically, we used the statement 
\verb|Comment.HAS_CREATOR| which always has out-degree of 1, the statement \verb|Forum.HAS_TAG|
which has an average out-degree of 3.5, and lastly the statement \verb|Forum.HAS_MEMBER| which has an 
average out-degree of 28 members.

\begin{table}
    \caption{Index scan rates (tuples read M/sec)}
    \vspace{-5pt}
    \label{table:scan1}
    \begin{tabular}{llrr}
    \toprule
    statement & frontier size & time (ms) & tpls (M/s)\\
    \midrule
    \verb|Comment.HAS_CREATOR| & 100,000 & 120.75 & 0.82 \\
    \verb|Comment.HAS_CREATOR| & 500,000 & 529.64 & 0.94 \\
    \verb|Forum.HAS_TAG| & 100,000 & 339.21 & 1.10\\
    \verb|Forum.HAS_TAG| & 500,000 & 1626.19 & 1.14\\
    \verb|Forum.HAS_MEMBERS| & 100,000 & 599.20 & 4.79 \\
    \verb|Forum.HAS_MEMBERS| & 500,000 & 2833.18 & 5.09 \\
    \bottomrule
    \end{tabular}
\end{table}

Table~\ref{table:scan1} lists the tuples retrieved by probing the index for different size of frontiers
(i.e., 100,000 and 500,000 frontier sizes). When the out-degree of the queried statement increases then we
notice an increase of the tuples retrieved up to 5 million tuples per second.
This is observed because for each index probe we read more tuples from the adjacency list.

\begin{table}
    \vspace{-5pt}
    \caption{Relation scan rates (tuples read M/sec)}
    \vspace{-5pt}
    \label{table:scan2}
    \begin{tabular}{llrr}
    \toprule
    statement & frontier size & time (ms) & tpls (M/s)\\
    \midrule
    \verb|Comment.HAS_CREATOR| & 100,000 & 1446.2 & 452.8  \\
    \verb|Comment.HAS_CREATOR| & 500,000 & 2116.05 & 312.5\\
    \verb|Forum.HAS_TAG| & 100,000 & 1645.35 & 407.2 \\
    \verb|Forum.HAS_TAG| & 500,000 & 2766.54 & 266.5\\
    \verb|Forum.HAS_MEMBERS| & 100,000 & 1779.04 & 376.6 \\
    \verb|Forum.HAS_MEMBERS| & 500,000 & 3258.59 & 226.3 \\
    \bottomrule
    \end{tabular}
\end{table}

Table~\ref{table:scan2} lists the tuples read by the relation scan operation for different sized input 
frontiers. Although relation scans can achieve maximum memory throughput (more than 500M tuples read per
second) different factors come to play. Large sized frontiers entail more costs for matching tuples to
input vertices (through hash probing of the frontier) and also more tuples are materialized per page 
scanned. Thus we can see a drop of tuples read when both frontier and out-degree increase.

\begin{figure*}
    \includegraphics[width=0.99\textwidth]{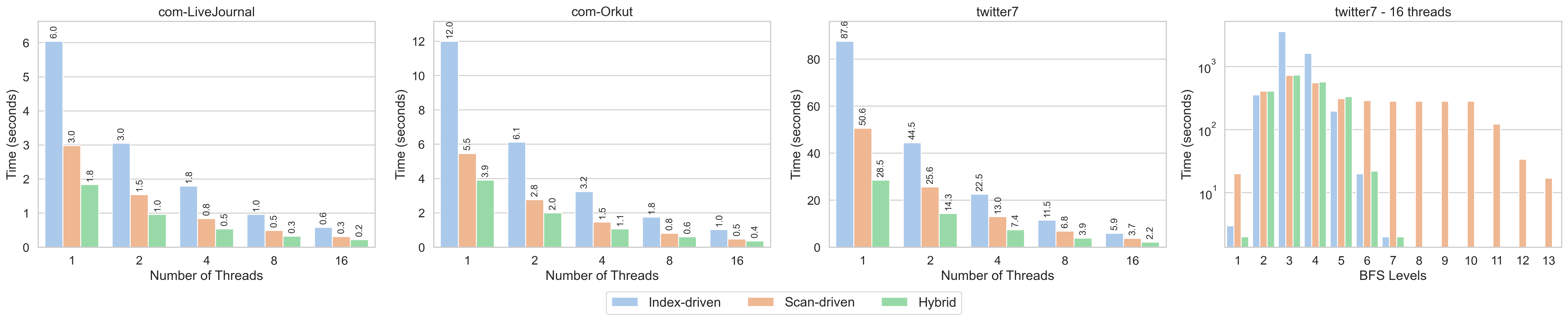}
    \vspace{-8pt}
    \caption{Comparison of index-based, scan-based and hybrid (default) BFS performance.}
    \vspace{-8pt}
    \label{fig:perf-bfs}
    \Description{BFS}
\end{figure*}

\paragraph{Graph Algorithms}
\label{sec:perf-graph}
Figure~\ref{fig:perf-bfs} displays a comparison of index-based, scan-based and hybrid BFS implementation, using three popular datasets from SuiteSparse\footnote{SuiteSparse: \url{https://suitesparse-collection-website.herokuapp.com/}}.
The characteristics of the three datasets used in this experiment and a fourth one, RMAT Scale-26, are displayed in Table~\ref{tab:prop}.
We carried out these experiments on an R6i instance using the \Peight storage engine without the higher level components to highlight trade-offs in the actual algorithm execution.
As described in Section~\ref{sec:rel}, SPOI tuples are stored in heap -- using scans of the heap in early iterations of algorithms such as BFS reads too much data.  
Index-driven scans use a secondary S (or O) index to find the SPOIs locations for specific S (or O), hence they have additional indirection, but are much more selective.
Hence especially when frontier is large, such as in levels 3-4 in most power-law graphs, this causes too many unnecessary memory indirections,
as seen in the last chart which presents the breakdown of execution time to levels in 16-thread parallel BFS on twitter7 dataset.
Our hybrid approach switches dynamically between index-driven and scan-driven access to achieve the best performance for all datasets in all thread counts.

\begin{table}
\caption{Properties of the test graph datasets.}
\vspace{-5pt}
\begin{tabular}{lrr}
    \toprule
    Dataset         & \# vertices   & \# directed edges \\
    \midrule
    com-LiveJournal	& 3,997,963	    &      69,362,378 \\
    com-Orkut	    &  3,072,441    &     234,370,166 \\
    twitter7    	& 41,652,230    &	1,468,364,884 \\
    \midrule
    RMAT Scale-26	& 67,108,864	&   2,147,483,648 \\
    \bottomrule
\end{tabular}
\label{tab:prop}
\end{table}

\begin{figure}[ht]
\vspace{-10pt}
    \centering
    \includegraphics[width=0.98\columnwidth]{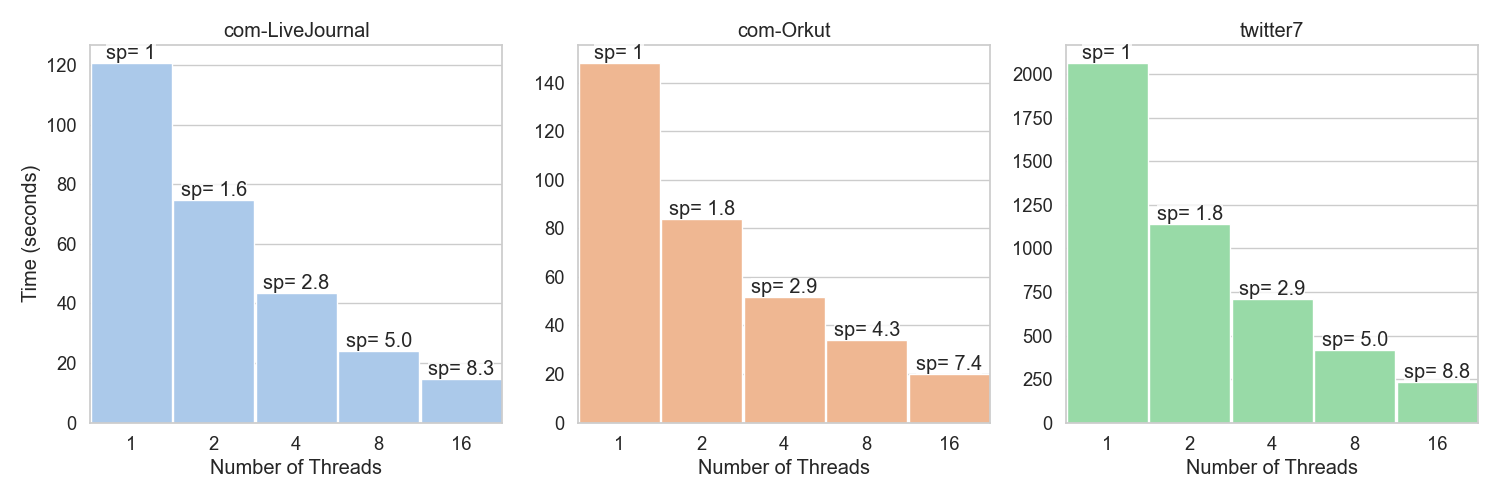}
    \vspace{-8pt}
    \caption{Closeness Centrality Performance on test datasets, using $n'=2,048$ source vertices. }
    \vspace{-8pt}
    \label{fig:perf-cc}
    \Description{CC}
\end{figure}

Figure~\ref{fig:perf-cc} displays the performance of the Closeness Centrality implementation. 
The complete (exact) centrality algorithm~\cite{Brandes2001} is equivalent to running one BFS from each vertex in the graph. 
Thus, for graph with $n$ vertices, $m$ edges, its runtime complexity is $O(n \cdot m)$, i.e., for each one of $n$ vertices, it requires $O(m)$ work for each BFS.
An approximated closeness centrality can be implemented by running BFS on a randomly selected {\em source vertices}, say with a cardinality of $n' \ll n$, hence achieving significant performance improvements~\cite{Madduri2009-IPDPS}.
The number of source vertices ($n'$) is an algorithm parameter that can be changed by the user.

We run the approximate algorithm by leveraging CPU-SpMM~\cite{Sariyuce15-JPDC}.
Furthermore to improve performance, CPU-SpMM concurrently runs $c$ concurrent BFSes using vectorized Sparse-Matrix Matrix multiplications, and repeats this process until all sampled vertices are used as a source for a BFS. 
The algorithm can be viewed as expanding the frontier of multiple BFSes level-by-level with each SpMM multiplication~\cite{Buluc11-cblas,Sariyuce14-MTAAP,Then2014-VLDB}. 
In the $\ell$-th multiplication, the sparse matrix represents the sparsity structure of graph, and the dense matrix can be viewed as concatenation of multiple columns, each representing the vertices in the current frontiers of one of the concurrent BFSes in the $\ell$-th level.
If the number of concurrent BFes ($c$) is small, let say in the extreme it is 1, and hence we use this algorithm to run single BFS, this algorithm will be less work efficient then the naive BFS algorithm.
For a graph with a diameter of $d$, its runtime is $O(d \cdot m)$ instead of just $O(m)$.
However, if $c > d$, which is true in many real-world graphs, this algorithm will be more work efficient.
Furthermore, since the access pattern of SpMM is more sequential then BFS, CPU-SpMM~\cite{Sariyuce15-JPDC} runs much faster than previous state-of-the-art CPU implementations. 
In our experiments we used $c=2,048$.
Our algorithm can heuristically choose a smaller value for $c$ (up to 32) when there is not enough memory to run the algorithm with larger $c$ values.

\begin{figure}[ht]
    \centering
    \includegraphics[width=0.98\columnwidth]{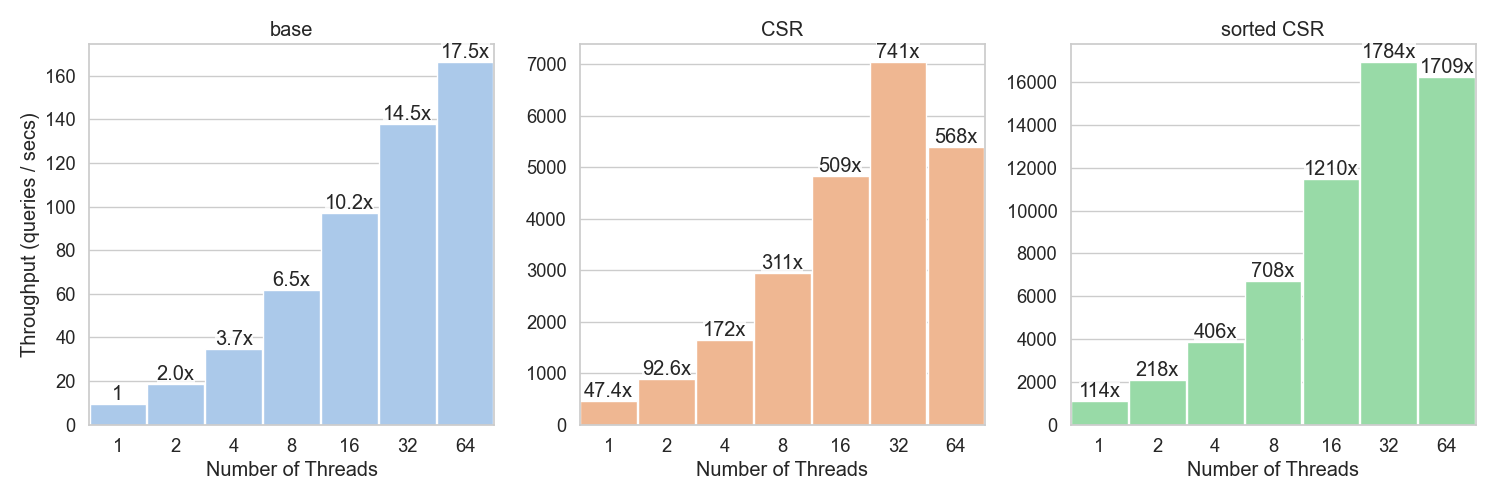}
    \vspace{-8pt}
    \caption{Performance of the EgoNet variants.}
    \label{fig:perf-egoNet}
    \vspace{-8pt}
    \Description{egoNet}
\end{figure}

Figure~\ref{fig:perf-egoNet} displays the performance of EgoNet algorithm on RMAT Scale-26.
The EgoNet algorithm finds the egocentric network of a given vertex to its $hopCount$ neighbors.
At each level, the algorithm includes top $k$ maximum (or minimum) weighted neighbors.
In this experiment, 128 EgoNet queries run with $k=100$ and $hopCount=2$ while varying the number threads used to concurrently run this experiment.
The first plot displays the performance of base algorithm running against dynamic live data.
This would require 2 level BFS-like traversal with filtering using index-driven traversal.
As seen in the figure, in this implementation throughput can be improved up to 17.5$\times$ using 32 threads.
The second plot displays the performance of the algorithm running on top of unsorted CSR.
Use of CSR, in particular, having edge and edge weight stored together in CSR, can improve the performance 47.4$\times$ on sequential execution and up to 741$\times$ using 32 threads.
Since the algorithm searches for top $k=100$ neighbors, if the CSR is generated by sorting the neighbors by weight (in any order),
algorithm can take advantage of this can look at either first or last $k$ edges in the adjacency list,
depending on sorting order.
The last plot displays the performance of the algorithm leveraging sorted CSR.
As seen in the figure, this gives additional 2.4$\times$ improvement over CSR, hence making improvement over base implementation up to 114$\times$ for sequential execution,
and up to 1784$\times$ improvement by using 32 threads.

